\documentclass[aps,reprint,longbibliography,superscriptaddress,showpacs]{revtex4-1}
\usepackage{CJK}
\usepackage{subfigure}
\usepackage{url}
\usepackage{amssymb}
\usepackage{amsmath}
\usepackage{dcolumn}
\usepackage{bm}
\usepackage{graphicx,float}
\usepackage{mathrsfs}
\usepackage[colorlinks,linkcolor=blue,anchorcolor=blue,citecolor=blue,urlcolor=blue]{hyperref}

\begin{document}
\begin{CJK}{GB}{}

\title{  Hybrid quantum system with  nitrogen-vacancy centers in diamond
 coupled to  surface phonon polaritons in  piezomagnetic superlattices }

\CJKfamily{gbsn}
\author{Peng-Bo Li (ÀîÅ)}

\affiliation {Shaanxi Province Key Laboratory of Quantum Information and Quantum Optoelectronic Devices,
 Department of Applied Physics, Xi'an Jiaotong University, Xi'an
710049, China}
\affiliation {Theoretical Quantum Physics Laboratory, RIKEN Cluster for Pioneering Research, Wako-shi, Saitama 351-0198, Japan}

%\affiliation {Center for Emergent Matter Science, RIKEN, Saitama 351-0198, Japan}

\author{Franco  Nori (Ò°Àí)}
\affiliation {Theoretical Quantum Physics Laboratory, RIKEN Cluster for Pioneering Research, Wako-shi, Saitama 351-0198, Japan}
\affiliation{Department of Physics, The University of Michigan, Ann Arbor, Michigan 48109-1040, USA}

\begin{abstract}
We investigate  a hybrid quantum system where an ensemble of nitrogen-vacancy (NV) centers in diamond is interfaced with  a piezomagnetic superlattice that supports surface phonon polaritons (SPhPs). We show that  the strong magnetic coupling between  the collective spin waves in the NV spin ensemble and the quantized SPhPs  can be realized, thanks to the subwavelength nature of the SPhPs and relatively long spin coherence times. The magnon-polariton coupling allows different modes of the SPhPs to be mapped and orthogonally stored in different spatial modes of excitation in the solid medium.  Because of its easy implementation and high tunability, the proposed  hybrid  structure with NV spins and piezoactive superlattices could be used  for quantum memory and quantum  computation.

\end{abstract}

\maketitle

\end{CJK}

\section{introduction}
Electron spins in solids are promising candidates for quantum memory and quantum computation because of their long coherence time and perfect compatibility with other solid state setups \cite{SCI-314-281,SCI-336-1280,nature-453-1043,natphys-2-408,Nat-Mater-11-143,NC-4-1743,prl-105-140501,PR-528-1,prx-7-031002}.  In  hybrid quantum systems, coherent coupling between ensembles of nitrogen-vacancy (NV) centers in diamond and superconducting quantum circuits has been demonstrated \cite{prl-105-140502,prl-105-210501,prl-107-060502,natphys-10-720,nature-478-221}. To further explore the potential of spin ensembles, spatial modes of collective spin excitations  can be used  to encode a register of qubits \cite{prl-103-070502,prl-105-140503,prx-4-021049}, which allows to implement  holographic quantum computing \cite{prl-101-040501}.  However, for superconducting quantum circuits, due to the long wavelength nature of microwave fields, they can only directly  couple to collective spin excitations with a vanishing phase variation. This directly limits the ability to
process quantum information in a spin ensemble via microwave photons. To make the best use of holographic techniques in spin ensembles, it is very appealing to coherently couple single microwave photons with collective spin wave excitations. This is quite challenging  in current experiments involving superconducting cavities, since the free-wavelength of microwave photons is always much larger than the dimensions of spin ensembles.
Here, we propose a novel protocol for this problem by coupling an ensemble of NV spins in diamond to surface phonon polaritons (SPhPs) in piezomagnetic superlattices.

SPhPs are electromagnetic surface modes resulting from the coupling between crystal vibrations and electromagnetic fields \cite{jpc-6-1266,jpc-7-3547,jvst-11-1004,ap-16-211,rpp-37-817}.
Analogous to surface plasmon polaritons  \cite{rpp-70-1,nature-424-824,prl-97-053002,prl-101-190504,nature-450-7168,prl-93-036404,prl-103-053602,prl-106-096801,prl-108-066401,rpp-78-013901,Nat-Photon-4-112,prl-110-126801},  SPhPs tightly bound to the  surface of a  dielectric material  often have a subwavelength confinement \cite{SCI-343-1125,NC-5-5221,NC-5-4782}. A piezoactive superlattice is a type of ordered microstructures, where the piezoelectric or piezomagnetic coefficient is periodically modulated  \cite{SCI-284-1822,prl-90-053903,jap-109-064110,prb-71-125106,apl-101-151109}.  SPhPs in piezoactive superlattices  can be tailored with engineered frequencies and bandwidths via a suitable
design of the superlattice \cite{apl-101-151109,prl-117-103201}. This opens the possibility for  generating SPhPs of microwave frequencies that can interact with  ensembles of NV spins.

We  show that there indeed exist SPhPs confined near the surface of a piezomagnetic superlattice formed by alternating layers of piezomagnetic materials. Then we provide a full quantum theory to describe the magnetic coupling between the collective spin wave excitations of the NV spin ensemble and the quantized SPhP modes  in the piezomagnetic superlattice.
When taking into account the dissipations of SPhPs and NV centers, coherent couplings can dominate the interactions and the strong coupling regime can be realized.
The achieved coupling strength can be  the same order of magnitude as that associated with superconducting quantum circuits \cite{prl-105-140502,prl-105-210501,prl-107-060502,natphys-10-720,nature-478-221}.
Unlike  earlier work  employing  superconducting qubits
or resonators, this hybrid structure exploits the magnon-SPhP  coupling, and
takes advantage of the subwavelength nature of SPhPs
and the excellent tunability and scalability of piezoactive
superlattices \cite{SCI-284-1822,prl-90-053903,jap-109-064110,prb-71-125106,apl-101-151109}.
This strong, and tunable magnon-polariton
coupling allows to implement the holographic techniques
with this hybrid structure \cite{prl-103-070502,prl-105-140503,prx-4-021049}.
The combined NV spins and piezomagnetic superlattices approach opens new routes for constructing novel hybrid quantum devices \cite{RMP-1,prl-103-043603,prl-117-015502, prappl-4-044003,np-2-636,prl-97-033003,prl-102-083602,pra-88-012329,prb-87-144516,npj-QI-3-28} with  solid state artificial structures, and could have wide
applications in a range of fields: from nanophotonics \cite{caldwell2015,Nat-Photon-11-149,nl-13-3690,ACS-photon-1-718,Nat-Photon-9-320} to  quantum information processing \cite{SCI-334-463,natphys-9-329,Nat-Photon-11-398,rp-1-120}.

\begin{figure}[t]
\centering
\centerline{\includegraphics[bb=36 0  789 215,totalheight=1.0in,clip]{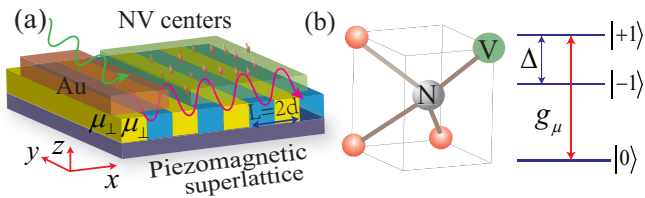}}
\caption{(Color online) (a) Schematic of an ensemble of NV centers in a diamond crystal located on the
surface of a piezomagnetic superlattice with a period $L=2d$, and dimensions $50\times 2\times 0.5 ~\text{mm}^3$.
The surface lies in the $xy$ plane, while the
SPhPs  propagate along the $x$ direction, where $x,y$, and $z$ are the principle axes of the piezomagnetic
crystal. Because a uniaxial material is used, we assume the $z$-axis to be the optical axis, which defines
$\mu^{(x)}=\mu^{(y)}=\mu_{\bot}$, and $\mu^{(z)}=\mu_{\parallel}$. (b) Schematic of an NV center with its vacancy (V) and nitrogen
atom (N), as well as three neighboring carbon atoms (left); Energy level diagram of the NV center (right).}
\end{figure}

\section{The setup }
As sketched in Fig.~1(a), an ensemble of NV centers in a diamond crystal is positioned above the surface of a semi-infinite periodic structure
composed of alternating layers of piezomagnetic materials  such as Terfenol-D or $\text{CoFe}_2\text{O}_4$. This kind of periodic artificial structure, with the piezomagnetic coefficient being periodically modulated, forms the so-called piezomagnetic superlattice \cite{prb-71-125106}. In this setup, a negative permeability can be realized, and a type of phonon polaritons typically bound to the interface between the surrounding medium and the piezomagnetic material can be created \footnote{See Appendixes for more
details}. A semi-infinite gold (Au) film is deposited on top of the superlattice, which is used as a broadband antenna for converting the
incident microwave field into strongly confined near fields at the gold edge \cite{SCI-343-1125,NatPhoton-9-674}.

We consider a piezomagnetic superlattice formed by Terfenol-D with a period of 1$\mu$m.  Figures~2(a,b)  display the calculated
effective permeability $\mu_{\perp}$ and $\mu_{\parallel}$ of the piezomagnetic superlattice \cite{Note1}.  As can be seen, for frequencies $\omega_{\perp L} <\omega<\omega_{\perp o}$, the effective permeability $\mu_{\perp}$  is negative, while $\mu_{\parallel}$ is positive.
The permeability tensor is invariant with respect to rotations about the $z$ axis.
If we then consider the frequency of a surface polariton of propagation vector $\vec{k}_p$, the presence of
the rotational symmetry of the permeability tensor means that the frequency must be independent of the orientation of the  in-plane wavevector $\vec{k}_p$ relative to the $x$ and $y$ axes \cite{rpp-37-817,jpc-7-3547}. Therefore, the surface polariton dispersion relation is independent of the direction of $\vec{k}_p$.

To derive the form of the dispersion relation, with no loss of generality, we may
assume that the propagation vector $\vec{k}_p$ lies along the $x$ direction \cite{rpp-37-817,jpc-7-3547}.
Figures~2 (c,d) show the dispersion relation for SPhPs propagating along the interface between vacuum and the piezomagnetic superlattice \cite{Note1}. We find that, in the spectral gap, the in-plane wavevector  is purely real, while the wave vector normal to the surface is purely imaginary \cite{NC-5-3300,SCI-348-1448}.  Thus, the fields remain localized and only propagate along the  interface.
Based on the results in Fig.~2(d), we estimate that the wavelength for the SPhP of frequency $\omega\sim 2\pi\times  3.4$ GHz is about $\lambda_p\sim 6$ mm, which is about one order smaller than  the free-space wavelength ($\lambda_0\sim 9$ cm) or that of microwave photons in a superconducting cavity.
 In this case, the effective volume of the electromagnetic fields can be significantly reduced near the SPhP resonance, which leads to a strong field enhancement.
\begin{figure}[b]
\centering
\centerline{\includegraphics[ bb=12 176 281 377,totalheight=2.65in,clip]{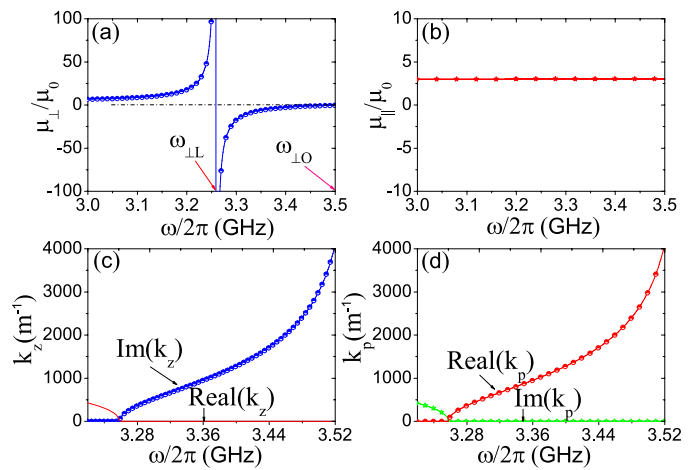}}
\caption{(Color online) (a,b): Calculated  effective permeability in the piezomagnetic superlattice formed by Terfenol-D with a period of 1 $\mu$m.  (c,d):  Dispersion relation of SPhPs on a plane interface between vacuum and the piezomagnetic superlattice. }
\end{figure}

We assume the SPhPs propagate along the $x$ direction with surface normal along the $z$ axis. The SPhPs can be approximated as TE fields with the magnetic field in the $x$-$z$ plane. In this case, the quantized  magnetic field  of the SPhP with mode $\vec{k}$  becomes \cite{Note1}
\begin{eqnarray}\label{B}
\vec{B}_{\vec{k}}&=&\sqrt{\frac{\hbar\omega(\vec{k})\mu_0}{2S}} \vec{b}_{\vec{k}}(z)\hat{a}_{\vec{k}}e^{ik_p x-i\omega(\vec{k}) t}+\text{H.c.},
\end{eqnarray}
where $\omega(\vec{k})$ is the frequency, $S$ is the quantization area,   and $\hat{a}_{\vec{k}}$ is the destruction operator for mode $\vec{k}$.
The polarization vector $\vec{b}_{\vec{k}}(z)$ is given by  \cite{Note1}
\begin{eqnarray}
\vec{b}_{\vec{k}}(z)&=& \mathcal {L}^{-1/2}(\vec{k})e^{-\text{Im}(k_z)z}(\vec{e}_x-\frac{k_p}{k_z}\vec{e}_z).
\end{eqnarray}
Here $\mathcal {L}(\vec{k})$ is the effective length of the mode, which depends on the geometry and magnetic response of the superlattice, and $\vec{e}_x$ and $\vec{e}_z$ are the unit vectors in the $x$ and $z$ directions.

\section{Coupling SPhP modes to spin waves}
We now proceed to consider the coupling between  NV spins and SPhPs. NV centers in diamond consist of a substitutional
nitrogen atom and an adjacent vacancy, which have a spin $S=1$
ground state, with zero-field splitting $D=2\pi\times 2.87$ GHz,
between the $\vert m_s=\pm1\rangle$ and $\vert m_s=0\rangle$ states.
For moderate applied magnetic fields ($B_z$  about several mT, and  compatible with the SPhP modes), which cause Zeeman
shifts of the states $\vert m_s=\pm1\rangle$,
one of the spin transitions of the NV center can be tuned into resonance with the SPhPs
mode. This allows us to isolate a two-level subsystem comprised by $\vert m_s=0\rangle$ and $\vert m_s=+1\rangle$, as shown in Fig. 1(b).

The interaction
of a single NV center located at $\vec{r}_0$
with  the  total magnetic field can be written as
\begin{eqnarray}
\hat{H}_\text{NV}=\hbar D \hat{S}{_{z}^{2}}+\mu_Bg_s\ B_z\hat{S}{_{z}}+\mu_Bg_s \vec{B}_{\vec{k}}(\vec{r}_0)\cdot\hat{\vec{S}},
\end{eqnarray}
with $g_s=2$ the Land\'{e} factor of the NV center,  $\mu_B$ the Bohr magneton,  and $\hat{\vec{S}}$ the spin operator of the NV center. Under the condition $|\Delta/2+D-\omega(\vec{k})|\ll\Delta/2$, with $\Delta=2\mu_Bg_sB_z/\hbar$, we can neglect the state $\vert m_s=-1\rangle$, due to the
external field moving it far out of resonance.
Then we can derive  the following Hamiltonian that describes the
interaction between a single NV spin and a  SPhP mode $\vec{k}$ \cite{Note1}
\begin{eqnarray}
\hat{\mathcal{H}}_{\vec{k}}&=&\frac{1}{2}\hbar \omega_0\hat{ \sigma}_{z}+\hbar \omega(\vec{k})\hat{a}_{\vec{k}}^\dag\hat{a}_{\vec{k}}\nonumber\\
&&+\frac{\hbar g_{\mu}(\vec{k},z_0)}{\sqrt{S}}  \hat{\sigma}_{+} \hat{a}_{\vec{k}}e^{ik_p x_0}+\text{H.c.},
\end{eqnarray}
with $\hat{\sigma}_z=\vert +1\rangle\langle +1\vert-\vert 0\rangle\langle 0\vert$, $\hat{\sigma}_+=\vert +1\rangle\langle 0\vert $, $\hbar\omega_0=\hbar D+\mu_Bg_sB_z$, and
\begin{eqnarray}
  g_{\mu}(\vec{k},z_0) &=& \frac{\mu_Bg_s}{2}\sqrt{\frac{ \omega(\vec{k})\mu_0}{\hbar \mathcal {L}(\vec{k})}}e^{-\text{Im}(k_z)z_0}.
\end{eqnarray}

\begin{figure}[t]
\label{Fig3}
\centering
\centerline{\includegraphics[bb= 193  404  362 472, totalheight=1.4in,clip]{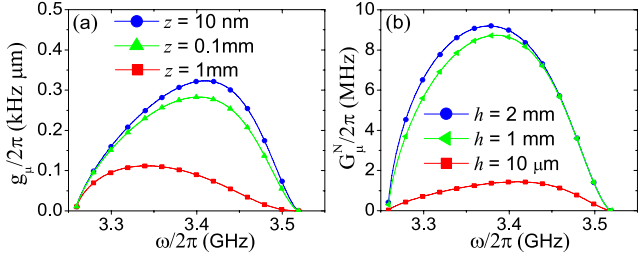}}
\caption{(Color online) (a) Coupling constant $g_{\mu}(\vec{k,}z) $ of a single NV spin for different positions $z$ interacting with a SPhP mode with the wave-vector $\vec{k}$.
(b)  Collective coupling constant $G_{\mu}^N(\vec{k})$  of an NV center ensemble in a diamond crystal with dimensions $20\times2\times h$ $\text{mm}^3$  and a density of $n\sim2\times10^{6}\mu \text{m}^{-3}$  (about 6 ppm) \cite{prl-105-140502,prl-107-060502} for different values of the thickness $h$.  Other parameters are chosen as those in Fig. 2.  }
\end{figure}
We now consider the coupling between the NV spin ensemble  and the SPhP modes.
As depicted in Fig.~1(a),  an ensemble of  NV
centers is doped into a  diamond crystal of thickness $h$, and located at positions $\vec{r}_i$, each of which with a fixed quantization axis pointing
along one of the four possible crystallographic directions. If the orientations are equally distributed among the
four possibilities, and the external field is homogeneous, then a quarter of the NV spins can be made resonant with the SPhP mode.
In such a case, we  have the following Hamiltonian for $N$ NV spins in the resonant
subensemble interacting with the quantized surface mode $\vec{k}$
\begin{eqnarray}\label{HN}
\hat{\mathcal{H}}^N_{\vec{k}}&=& \sum_{i=1}^{N}\frac{1}{2}\hbar\omega_{i}\hat{ \sigma}_{z}^{i}+\hbar \omega(\vec{k})\hat{a}_{\vec{k}}^\dag\hat{a}_{\vec{k}}\nonumber\\&&+\sum_{i=1}^{N}\frac{\hbar g_{\mu}(\vec{k},z_i)}{\sqrt{S}}(  \hat{\sigma}_{+}^{i} \hat{a}_{\vec{k}}e^{ik_p x_i}+\text{H.c.}),
\end{eqnarray}
where $\omega_{i}=\omega_{0}+\delta_{i}$, and $\delta_{i}$  are random offsets accounting for the inhomogeneous broadening of the spin ensemble.  These inhomogeneous broadening terms are usually  on the order of megahertzs, whose effect can be described  by spin dephasing.

To further simplify the model, we introduce the collective operators for the spin wave modes in the NV  ensemble
\begin{equation}
\hat{S}_{\vec{k}}^\dag=\frac{1}{\sqrt{N}g^N_{\mu}(\vec{k})}\sum_{i=1}^{N}g_{\mu}(\vec{k},z_i)\hat{\sigma}_{+}^{i}e^{ik_p x_i},
\end{equation}
 with $g^N_{\mu}(\vec{k})=\sqrt{ \sum_{i=1}^{N}|g_{\mu}(\vec{k},z_i)|^2/N}$.
These spin wave modes are orthogonal if the size of the diamond is much larger than  the wavelength of the SPhP modes, and the separation
between the NV spins is smaller than the SPhP wavelength. We consider the commutator $[\hat{S}_{\vec{k}_i}, \hat{S}_{\vec{k}_j}^\dag]\equiv D(\vec{k}_j-\vec{k}_i)$ in the fully polarized limit. We find that
\begin{equation}
D(\vec{k}_j-\vec{k}_i)\sim \frac{1}{N}\int^{l/2}_{-l/2}e^{-i(k_{p,i}-k_{p,j})x}dx,
\end{equation}
 with $l$ the extent of the sample along the $x$ direction. When $\Delta k=k_{p,i}-k_{p,j}=2\pi/l$, the mode overlap $D(\vec{k}_j-\vec{k}_i)=0$, which means
the spin wave modes  in the strongly polarized limit are orthogonal.

Then we have the following effective interaction Hamiltonian between the SPhP mode $\hat{a}_{\vec{k}}$ and the spin wave $\hat{S}_{\vec{k}}$
\begin{eqnarray}\label{HN2}
\hat{\mathcal{H}}^I_{\vec{k}}&=& \hbar G_{\mu}^N(\vec{k})(  \hat{S}^\dag_{\vec{k}}\hat{a}_{\vec{k}} +\text{H.c.}).
\end{eqnarray}
Here the collective coupling strength is given by
\begin{eqnarray}
G_{\mu}^N(\vec{k})&=&g_{\mu}^N(\vec{k})\sqrt{\frac{N}{S}}
=\sqrt{\frac{1}{S}\sum_{i=1}^{N}|g_{\mu}(\vec{k},z_i)|^2}\nonumber\\
&=& \frac{1}{2}\sqrt{ n \int_{0}^{h}|g_{\mu}(\vec{k},z)|^2 dz}
\end{eqnarray}
where we assume a continuum of layers in the $z$ direction with a thickness $h$, and a volume density of NV spins $n=4N/(Sh)$. Obviously, the effective coupling  strength
between the NV spins and the quantized surface field is enhanced by a factor of $\sqrt{n}$. According to Eq. (\ref{HN2}) the SPhP mode
and the spin wave mode behave as two coupled oscillators with a coupling strength $G_{\mu}^N(\vec{k})$. This allows any states of the two systems to be interchangeably mapped between them.

It is very useful to compare the coupling between NV spins and SPhPs with that between spin ensembles and other setups \cite{prl-105-140502,prl-105-210501,prl-107-060502,nature-478-221}.
In this hybrid system, the surface polariton modes are coupled to collective electron spin wave excitations in the spin ensemble, in direct contrast with other setups involving superconducting qubits or resonators \cite{prl-105-140502,prl-105-210501,prl-107-060502,nature-478-221}. For the latter, only the collective spin excitations with a vanishing phase variation are employed, because of the long wavelength nature of microwave fields. However, in our proposed device,
even though it works in the microwave  range, the spatial modes of spin excitations must be considered, due to the subwavelength nature of the SPhPs. This will be very useful for holographic quantum computation with spin ensembles, which employs collective spin wave excitations to encode a register of qubits \cite{prl-103-070502,prl-105-140503,prx-4-021049,prl-101-040501}.

In Fig.~3, we plot the calculated single-spin coupling constant $g_{\mu}(\vec{k,}z) $ and the collective coupling $G_{\mu}^N(\vec{k})$  as a function
of the frequency within the negative gap. From Fig.~3(a), we find that obviously $g_{\mu}$ varies with the distance $z$ and decreases as $z$ increases, which indicates that the SPhP modes decay exponentially along the direction normal to the interface. Figure~3(b)  shows that the collective coupling strength $G_{\mu}^N(\vec{k})$ depends strongly on the thickness $h$  but tends to saturate for thick enough crystals.  When $\omega\sim 2\pi\times 3.4$ GHz, the collective coupling strength can reach $2\pi\times 9$ MHz.  This coupling strength is comparable to
that of an NV spin ensemble  coupled to a superconducting flux qubit \cite{nature-478-221,prl-105-210501} or a coplanar waveguide cavity \cite{prl-105-140502,prl-107-060502}.

In actual crystals, due to scattering, absorption, and other process, the decay of the SPhP mode ($\kappa_\text{SPhp}$) should be taken into consideration \cite{Note1}. It has been shown \cite{jpc-7-3547} that  the decay of the SPhP mode is frequency dependent, and near the SPhP resonance frequency the SPhP decay is approximately equal to the damping constant of the crystal. For piezomagnetic crystals like Terfenol-D or $\text{CoFe}_2\text{O}_4$, the damping constant can be approximated as $\Gamma \sim 0.001\omega_0$ \cite{jap-109-064110}, with $\omega_0$ the resonance frequency for the SPhP mode.  In this case, we can estimate the
decay of the SPhP mode as $\kappa_\text{SPhp}\sim 2\pi \times  3.4$ MHz.
To further increase the Q-factor of the surface modes so that the
strong-coupling regime can be entered more easily, two main
strategies can been pursued. The first concentrates on reducing the
damping of the material. The second  exploits cavities incorporated into superlattice structures \cite{natphys-9-329,nl-13-3690,ACS-photon-1-718},
which can combine the benefits of a high Q-factor and small mode volume.

For NV spins,
the $T_2$  time for a spin ensemble may be reduced to $T_2^*$  because of  interactions with nearby lattice nuclei and
paramagnetic impurities \cite{Note1}.
The hyperfine interaction with lattice $^{13}\text{C}$ will be
detrimental to the electron spin coherence, but this can be overcome when
using isotopically purified $^{12}$C diamond. For $^{14}$N nuclear spin,  due to its the long relaxation time,
the nuclear spin does not lead to decoherence but  can actually be used as a
resource for quantum memory.
Recent experiments  demonstrate that the dephasing time for an NV spin ensemble is still in the microsecond range, with
$\gamma_s\sim 2\pi\times 3$ MHz \cite{prl-105-140502}.  If other methods, such as the spin echo techniques, are used \cite{nature-461-1265}, then the spin dephasing time will be extended from $T_2^*$ to $T_2$, which is close to the intrinsic spin
coherence time (on the order of kHz).

A useful  measure of the coupling efficiency is
the cooperativity $C=(G_{\mu}^N)^2/(\gamma_s\kappa_\text{SPhP})$. The strong coupling regime can be reached if the collective coupling strength $G_{\mu}^N(\vec{k})$ exceeds
both the electronic spin dephasing rate $\gamma_s$  and the
intrinsic damping rate of the SPhPs mode $\kappa_\text{SPhP}$, $G_{\mu}^N(\vec{k})> \{ \gamma_s,\kappa_\text{SPhP}\}$, i.e., $C>1$.
Figure 4(a) shows the cooperativity decreasing as the superlattice period $L=2d$ increases.
In our hybrid magnon-polariton system, we choose $L=2d=1\mu \text{m}$, and obtain $C\sim 90$ with a moderate spin dephasing
rate.
Moreover,  in Fig.~4(b) we plot the eigenfrequencies $\omega_{\pm}$ of the coupled magnon-polariton system as the spin transition frequency is varied through resonance with the
SPhP mode by changing the Zeeman splitting.  The avoided crossing with a clear
splitting shows that the magnon mode strongly couples to the SPhP mode.

\begin{figure}[t]
\label{Fig4}
\centering
\centerline{\includegraphics[bb=25 431 333 568,totalheight=1.4in,clip]{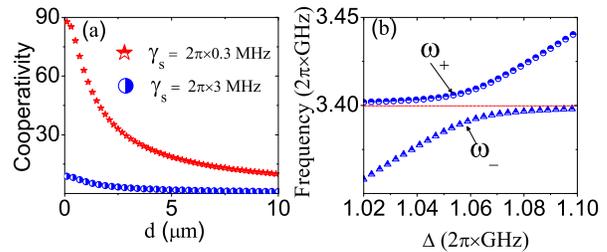}}
\caption{(Color online)  (a) Cooperativity $C$ as a function of the superlattice period $L=2d$. Here
$\kappa_{\text{SPhP}}\sim 0.001 \omega_{\perp L}$, and $h=1$ mm. Other parameters are chosen the same as those in Fig.~3. (b)  Eigenfrequencies of the coupled magnon-polariton system. Here $G_{\mu}^N=2\pi\times 9$ MHz, $\gamma_s\sim 0.2~G_{\mu}^N$, and $\kappa_{\text{SPhP}}\sim 0.3~G_{\mu}^N$. }
\end{figure}

We now consider the preparation  and detection of the relevant  states of the hybrid system.
NV centers can be prepared in their ground spin state $\vert m_s=0\rangle$ by using the spin selective optical
pumping method \cite{prx-4-021049}. As for SPhPs,
it is possible to excite a SPhP mode with a fixed wave vector
by applying the edge-launching or edge-coupling method \cite{SCI-343-1125,apl-92-203104}.
Actually, a recent experiment \cite{NatPhoton-9-674},  has demonstrated that SPhPs with the plane wave propagation can be
launched at the edge of a semi-infinite gold film deposited on top of a hexagonal boron nitride slab.
Therefore, we may envision that in principle the same approach can be used in our setup to excite a plane wave SPhP mode
in a piezomagnetic superlattice. In this way, the SPhP mode with the propagating wave vector $\vec{k}_p$ can produce a continuous wave magnetic field near the surface, which will strongly couple to the collective spin wave mode described by $\hat{S}_{\vec{k}}$.
Then, the state of a  SPhP mode will be mapped to a specific spin wave. Excitations stored in this way may be
detected by applying a gradient pulse that converts a particular spin wave back into a uniform transverse magnetization \cite{prl-103-070502,prl-105-140503,prx-4-021049}.
As demonstrated in recent experiments \cite{ prx-7-031002,eprint-180207100},  a direct measurement of the spin polarization can be realized by optically detected magnetic resonance.

\section{Storage of different SPhP modes}
Since the $k$ modes of the spin ensemble in the strongly polarized limit behave as a large number of independent harmonic oscillators,
it is possible to orthogonally store multiple SPhP modes in these spin wave modes with the magnon-polariton coupling. We consider the storage of
two different SPhP modes with the propagation vectors $k_1$ and $k_2$, and the polariton operators $\hat{a}_{\vec{k}_1}$  and $\hat{a}_{\vec{k}_2}$. As described above, the propagation vectors should satisfy $\vert k_1-k_2\vert=2\pi/l\sim 314$ to ensure that the corresponding spin wave modes are orthogonal. Based on the results given in Fig.~2(b) and Fig.~3(b), we  find that these modes exist and
can strongly couple to the spin waves. So if the SPhP modes $k_1$ and $k_2$ are excited by, for instance, the edge-launching method, then with
the interaction described by Eq. (\ref{HN2}), we can map the states of the SPhP modes into the collective spin wave modes  $\hat{S}_{\vec{k}_1}$  and $\hat{S}_{\vec{k}_2}$, through the swap gate $\hat{U}_\text{sw}=e^{-i \hat{\mathcal{H}}^I_{\vec{k}}T_{\pi/2}/\hbar}$, with $T_{\pi/2}=\pi/(2G_{\mu}^N)$. In principle, the above procedure  can
apply to the general  multimode storage case.

\section{Conclusions}
We have proposed a hybrid quantum device where an ensemble of NV centers in a diamond crystal is interfaced with a piezomagnetic superlattice supporting SPhPs. We have shown that the magnetic coupling between the collective spin waves and SPhPs can be tailored through a suitable design of the piezomagnetic superlattice. This strong  magnon-polariton coupling allows us to implement the holographic techniques with this hybrid device. Such a device could also be used as a coherent interface for other quantum systems such as superconducting qubits, cold atoms, and polar molecules.

\section*{acknowledgement}
P.B.L. acknowledges helpful discussions with  Peter Rabl, Carlos   S\'{a}nchez Mu\~{n}oz, Zhou Li, and Jiteng Sheng. P.B.L was supported by the NSFC under
 Grant Nos. 11774285 and 11474227. F.N. was  supported in part by the:
MURI Center for Dynamic Magneto-Optics via the
Air Force Office of Scientific Research (AFOSR) (FA9550-14-1-0040),
Army Research Office (ARO) (Grant No. 73315PH),
Asian Office of Aerospace Research and Development (AOARD) (Grant No. FA2386-18-1-4045),
Japan Science and Technology Agency (JST) (the ImPACT program and CREST Grant No. JPMJCR1676),
Japan Society for the Promotion of Science (JSPS) (JSPS-RFBR Grant No. 17-52-50023),
RIKEN-AIST Challenge Research Fund, and the
John Templeton Foundation.

\appendix

\section{Theory of surface phonon polaritons in piezomagnetic superlattices}
\subsection{Phonon polaritons in a piezomagnetic superlattice}
We study electromagnetic  waves propagating in a semi-infinite periodic structure composed of alternating layers of piezomagnetic materials such as Terfenol-D or $\text{CoFe}_2\text{O}_4$. This kind of periodic artificial structure, with the piezomagnetic coefficient being periodically modulated, forms the so-called piezomagnetic superlattice \cite{prb-71-125106}. In this artificial microstructure, the lattice vibrations will induce spin waves because of the piezomagnetic effect. Because the produced spin waves will in turn emit electromagnetic  waves  that interfere with the original electromagnetic
waves, the lattice vibration will couple strongly with the electromagnetic waves, leading to  polariton excitations. Near the piezomagnetic polariton resonance, an effective negative permeability can be implemented.

As displayed in Fig. 1 of the main text, the piezomagnetic superlattice is arranged along the $x$ axis, and we assume the transverse dimensions are very large compared with an acoustic wavelength so that a one dimensional model is valid.
The piezomagnetic equations describing the interaction between electromagnetic waves and acoustic waves are \cite{prb-71-125106,jsv-315-146,jsv-292-626}
\begin{eqnarray}
  T_{ij} &=& c_{ijkl}u_{kl}+q_{ijk} H_{k} \\
  B_i &=& \mu_{ik}^sH_k-q_{ikl}u_{kl}
\end{eqnarray}
where $T_{ij}, B_i,H_{k} $  are the stress tensor,  magnetic displacement, and magnetic field; $c_{ijkl},u_{kl},q_{ijk}$
are the elastic tensor, strain tensor, and piezomagnetic coefficient; $\mu_{ik}^s$ is the static magnetic permeability.
The piezomagnetic coefficient is periodically modulated with the form
\begin{eqnarray}
q(x)&=&\left\{
          \begin{array}{ll}
            +q, & \hbox{in positive domains \quad($0\leq x<d$);} \\
            -q, & \hbox{in negative domains \quad($d\leq x<2d$).}
          \end{array}
        \right.
\end{eqnarray}

From Newton's law $\rho \partial^2u_j/\partial t^2=(\partial/\partial x_i)T_{ij}$, we have
\begin{eqnarray}
 \rho \frac{\partial^2 u_j}{\partial t^2}=c_{ijkl}\frac{\partial^2 u_k}{\partial x_i\partial x_l}+\frac{\partial[q_{ijk}(x)H_k]}{\partial x_i},
\end{eqnarray}
with $u_j$ being the displacement along the Cartesian coordinate $x_j (x_1=x,x_2=y,x_3=z)$, and $\rho$ the mass density.
If the $x_1-x_2$ plane is taken as the isotropic plane of materials, the piezomagnetic tensor matrix can be written in the
following Voigt form \cite{prb-71-125106,jsv-315-146,jsv-292-626}
\begin{eqnarray}
q=\left( \begin{array}{cccccc}
                   0 & 0 & 0 & 0 & q_{15} & 0 \\
                   0 & 0 & 0 & q_{15} & 0 & 0 \\
                   q_{31}& q_{31} & q_{33} & 0 & 0 & 0 \\
                 \end{array}
                 \right)
\end{eqnarray}
and the elastic coefficient has the form \cite{prb-71-125106,jsv-315-146,jsv-292-626}
\begin{eqnarray}
 C=\left(
     \begin{array}{cccccc}
       c_{11} & c_{12} & c_{13} & 0& 0 & 0\\
       c_{12} & c_{11} & c_{13} & 0 & 0 & 0 \\
       c_{13}& c_{13} & c_{33} & 0 & 0 & 0 \\
       0 & 0 & 0 & c_{44} & 0 & 0 \\
       0 & 0 & 0 & 0 & c_{44} & 0 \\
       0& 0 & 0 & 0 & 0 & c_{66} \\
     \end{array}
   \right)
\end{eqnarray}
The static magnetic permeability tensor is
\begin{eqnarray}
\mu = \left(
        \begin{array}{ccc}
          \mu_{11}^s & 0 & 0 \\
          0 & \mu_{11}^s &0 \\
          0& 0 & \mu_{33}^s \\
        \end{array}
      \right)
\end{eqnarray}

With the above equations we can solve the piezomagnetic problem by using the Fourier transformation. First, we solve
a simple one dimensional model, which can be generalized to a more general case. The piezomagnetic equations pertaining to this
case are
\begin{eqnarray}
T_{11}&=&c_{11}u_{11}+q_{31}H_3=c_{11}\frac{\partial u_1}{\partial x}+q_{31}(x)H_3\\ \label{S1}
B_3&=& \mu_{11}^sH_3-q_{31}(x)u_{11}\\
 \rho \frac{\partial^2 u_1}{\partial t^2}&=&c_{11}\frac{\partial^2 u_1}{\partial x^2}+\frac{\partial[q_{31}(x)H_3]}{\partial x}.
\end{eqnarray}
By using the Fourier transformation
\begin{eqnarray}
u_1(x,t)&=&\int u(q)e^{i(\omega t-qx)}dq\nonumber\\
H_3(x,t)&=&\int H(k)e^{i(\omega t-kx)}dk\nonumber\\
q_{31}(x)&=&\sum_{m\neq0}\frac{i(1-\cos m\pi)}{m\pi}q_{31}e^{-iG_mx}\nonumber\\
&=& \sum_{m\neq0}F_mq_{31}e^{-iG_mx}, G_m=m\frac{\pi}{d},
\end{eqnarray}
we have
\begin{multline}\label{12}
  \int(\rho \omega^2-c_{11}q^2)u(q)e^{-qx}dq \\
=-\frac{\partial}{\partial x}\left[\sum_{m}F_mq_{31}e^{-iG_mx}\int H(k)e^{-ikx}dk\right] \\
=-\int \sum_{m}F_mq_{31}(-i)(k+G_m)H(k)e^{-i(k+G_m)x}dk.
\end{multline}

For photons with long wavelength $k\rightarrow0$ or $k\ll G_m$, Eq. (\ref{12}) becomes
\begin{eqnarray}
  \int(\rho \omega^2-c_{11}q^2)u(q)e^{-qx}dq&=&\sum_{m}iF_mq_{31}G_m H(k)e^{-i(k+G_m)x}\nonumber\\
\end{eqnarray}
In order to make the two sides be equal,  $q=k+G_m$ must be satisfied. Then we have
\begin{eqnarray}
u_1(q=k+G_m)=iF_mq_{31}\frac{G_m}{\rho \omega^2-c_{11}G_m^2}H(k)
\end{eqnarray}
and
\begin{eqnarray}\label{15}
u_{11}(x,t)&=&\int \left[F_mq_{31}\frac{G_m^2}{\rho \omega^2-c_{11}G_m^2}H(k)e^{i(\omega t-kx)}e^{-iG_mx}\right]dk\nonumber\\
&=&F_mq_{31}\frac{G_m^2}{\rho \omega^2-c_{11}G_m^2}e^{-iG_mx}H_3(x,t)\nonumber\\
&=&\varphi(x)H_3(x,t).
\end{eqnarray}
Substituting Eq. (\ref{15}) into Eq. (\ref{S1}), we have
\begin{eqnarray}
B_3&=&\mu_{11}^sH_3-q_{31}(x)\varphi(x)H_3(x,t)\nonumber\\
&=&\mu(x)\mu_0H_3(x,t).
\end{eqnarray}
For long wavelength $k\rightarrow0$, the piezomagnetic superlattice can be assumed to be homogeneous, and
the space average value of $\mu(x)$ should be used \cite{prl-90-053903}
\begin{eqnarray}
\tilde{\mu}_{\perp}(\omega)&=&\mu_{\perp}(\omega)\mu_0\nonumber\\
&=&\mu_{11}^s+\frac{1}{2d}\int^{2d}_0\mu(x)dx\nonumber\\
&=&\mu_{11}^s+\frac{q_{31}^2/d^2\rho }{\omega_{\perp L}^2-\omega^2}\nonumber\\
&=&\mu_{11}^s\frac{\omega_{\perp o}^2-\omega^2}{\omega_{\perp L}^2-\omega^2},
\end{eqnarray}
with
\begin{eqnarray}
\omega_{\perp L}^2&=&c_{11}\pi^2/\rho d^2\\
\omega_{\perp o}^2&=& \omega_{\perp L}^2+q_{31}^2/d^2\rho\mu_{11}^s.
\end{eqnarray}
Based on the same reasoning \cite{prb-71-125106}, we can obtain the following relations
\begin{eqnarray}
\mu_{\parallel}(\omega)&=&\mu_{33}^s/\mu_0\frac{\omega_{\parallel o}^2-\omega^2}{\omega_{\parallel L}^2-\omega^2}\\
\omega_{\parallel L}^2&=&c_{33}\pi^2/\rho d^2\\
\omega_{ \| o}^2&=& \omega_{\parallel L}^2+q_{33}^2/d^2\rho\mu_{33}^s.
\end{eqnarray}
Therefore, the effective magnetic permeability tensor is
\begin{eqnarray}
\mu(\omega) = \left(
        \begin{array}{ccc}
          \mu_{\perp}(\omega) & 0 & 0 \\
          0 & \mu_{\bot} (\omega)&0 \\
          0& 0 & \mu_{\parallel}(\omega) \\
        \end{array}
      \right).
\end{eqnarray}

\subsection{The surface phonon polariton dispersion relation}

We now proceed to discuss  the electromagnetic waves propagating at the boundary between vacuum ($\epsilon_1=1,\mu_1=1$) and the piezomagnetic superlattice ($\epsilon_2,\mu_{\perp},\mu_{\parallel}$).
The permeability tensor is invariant with respect to rotations about the $z$ axis.
If we then consider the frequency of a surface polariton of propagation vector $\vec{k}_p$, the presence of
the rotational symmetry of the permeability tensor means that the frequency must be independent of the orientation of the  in-plane wavevector $\vec{k}_p$ relative to the $x$ and $y$ axes \cite{rpp-37-817,jpc-7-3547}. Therefore, the surface polariton dispersion relation is independent of the direction of $\vec{k}_p$.
We consider the interface described in Fig. 1 of the main text, where the surface lies in the $xy$ plane.
To derive the form of the dispersion relation, with no loss of generality, we may
assume that the propagation vector $\vec{k}_p$ lies along the $x$ direction \cite{rpp-37-817,jpc-7-3547},
and look for TE wave solutions of Maxwell's equations in which the magnetic fields vary as the following form in both media \cite{rpp-37-817}:
\begin{eqnarray}
\vec{H}_1&=&\left(
              \begin{array}{c}
              H_{1x} \\
               0 \\
                 H_{1z}\\
              \end{array}
            \right)e^{ik_{p1}x-i\omega t} e^{i k_{1z}z}\\
\vec{H}_2&=&\left(
              \begin{array}{c}
              H_{2x} \\
               0 \\
                 H_{2z}\\
              \end{array}
            \right)e^{ik_{p2}x-i\omega t} e^{i k_{2z}z}
\end{eqnarray}
and the electric fields are in the $y$ direction:
\begin{eqnarray}
\vec{E}_1&=&\left(
              \begin{array}{c}
              0\\
               E_{1y}  \\
                 0\\
              \end{array}
            \right)e^{ik_{p1}x-i\omega t} e^{i k_{1z}z}\\
\vec{E}_2&=&\left(
              \begin{array}{c}
             0 \\
                E_{2y} \\
                 0\\
              \end{array}
            \right)e^{ik_{p2}x-i\omega t} e^{i k_{2z}z}.
\end{eqnarray}
The $\vec{k}$ vector parallel to the interface is conserved
\begin{eqnarray}\label{S27}
k_{p1}=k_{p2}=k_p.
\end{eqnarray}
From the boundary condition $\vec{n}\times (\vec{H}_2-\vec{H}_1)=0$,  we have
\begin{eqnarray}
H_{1x}=H_{2x}=H_x.
\end{eqnarray}
The wave numbers in medium 1 satisfy
\begin{eqnarray}
k_{p}^2+k_{1z}^2=\epsilon_1\mu_1k_0^2=\frac{\omega^2}{c^2}.
\end{eqnarray}
From $\nabla\cdot \vec{B}=0$, we have
\begin{eqnarray}
k_pH_{x}+k_{1z}H_{1z}=0.
\end{eqnarray}
In the magnetic superlattice, we have the following relations \cite{rpp-37-817}
\begin{eqnarray}
\frac{k_p^2}{\mu_{\|}}+\frac{k_{2z}^2}{\mu_{\perp}}=k_0^2\epsilon_2
\end{eqnarray}
\begin{eqnarray}\label{S32}
 k_p\mu_{\perp}H_x+k_{2z}\mu_{\parallel}H_{2z}=0.
\end{eqnarray}

With Eqs. (\ref{S27}) to (\ref{S32}), we can obtain the following relations for the wave numbers \cite{jpc-7-3547,prb-85-085442}
\begin{eqnarray}\label{S33}
k_p^2&=&\frac{(\mu_1\epsilon_2-\mu_{\perp})\mu_{\parallel}\mu_1}{\mu_1^2-\mu_{\perp}\mu_{\parallel}}k_0^2
\end{eqnarray}
\begin{eqnarray}
k_{1z}^2&=&\frac{\mu_1-\mu_{ \parallel}\epsilon_2}{\mu_1^2-\mu_{\perp}\mu_{\parallel}}k_0^2\mu_1^2
\end{eqnarray}
\begin{eqnarray}\label{S35}
k_{2z}^2&=&\frac{\mu_1-\mu_{ \parallel}\epsilon_2}{\mu_1^2-\mu_{\perp}\mu_{\parallel}}k_0^2\mu_{\perp}^2.
\end{eqnarray}
Based on Eqs. (\ref{S33})-(\ref{S35}), and using the material parameters given in table \ref{tab:table1} \cite{ijss-44-4593,jap-110-014508}, we can
investigate the dispersion relation of SPhPs in the piezomagnetic superlattice.
\begin{table}%The best place to locate the table environment is directly after its first reference in text
\caption{\label{tab:table1}%
Material parameters for Terfenol-D considered in this work (see \cite{ijss-44-4593,jap-110-014508}).
}
\begin{ruledtabular}
\begin{tabular}{llc}
Term &
Value  &
Units \\
\colrule
Mass density $\rho$ & $9.23\times10^3$ & kg/$\text{m}^3$\\
 $c_{11}$& $5.5\times10^{10}$ & N/$\text{m}^2$\\
$c_{33}$ &$5.5\times10^{10}$ & N/$\text{m}^2$\\
 $c_{44}$ & $1.2\times10^{10}$ & N/$\text{m}^2$\\
 $q_{13}$ & $-200$ & N/Am\\
$q_{33}$& 400& N/Am\\
$\mu_{11}^s$ & $6.23\times 10^{-6}$  &  $\text{N}^2$/A\\
$\mu_{33}^s$   & $6.23\times 10^{-6}$ &  $\text{N}^2$/A\\
$\epsilon_2$ & $10^3$ &---
\end{tabular}
\end{ruledtabular}
\end{table}
We find that there indeed exist SPhPs that propagate along the interface between vacuum and the  piezomagnetic superlattice
over a range of frequencies, within which the  permeability $\mu_{\perp}$ along
the propagation direction is negative, and $\mu_{\parallel}$ is positive, leading to the formation of a piezomagnetic superlattice.
The numerical results are given in
Fig.~2 in the main text.
Through Eqs. (\ref{S27})-(\ref{S35}), we can also obtain the relation between the field components
\begin{eqnarray}
 \frac{|H_x|^2}{|H_{1z}|}=-\frac{k_{1z}^2}{k_p^2}=-\frac{\mu_1(\mu_1-\mu_{\parallel}\epsilon_2)}{\mu_{\parallel}(\mu_1\epsilon_2-\mu_{\perp})}
\end{eqnarray}
\begin{eqnarray}
 \frac{|H_x|^2}{|H_{2z}|}=-\frac{\mu_{\parallel}^2k_{2z}^2}{\mu_{\perp}^2k_p^2}=-\frac{\mu_{\parallel}(\mu_1-\mu_{\parallel}\epsilon_2)}{\mu_1(\mu_1\epsilon_2-\mu_{\perp})}.
\end{eqnarray}
If we define
\begin{eqnarray}
 \vert H_x\vert^2+\vert H_{1z}\vert^2&=&\vert H_1\vert^2\\
  \vert H_x\vert^2+\vert H_{2z}\vert^2&=&\vert H_2\vert^2,
\end{eqnarray}
then we can obtain
\begin{eqnarray}
\vert H_x\vert^2&=&-\frac{\vert H_1\vert^2\mu_1(\mu_1-\mu_{\parallel}\epsilon_2)}{2\mu_1\mu_{\parallel}\epsilon_2-\mu_1^2-\mu_{\parallel}\mu_{\perp}}\nonumber\\
&=&-\frac{\vert H_2\vert^2\mu_{\parallel}(\mu_1-\mu_{\parallel}\epsilon_2)}{\mu_1(\mu_1\epsilon_2-\mu_{\perp})-\mu_{\parallel}(\mu_1-\mu_{\parallel}\epsilon_2)}
\end{eqnarray}
\begin{eqnarray}
\vert H_{1z}\vert^2=\frac{\vert H_1\vert^2\mu_{\parallel}(\mu_1\epsilon_2-\mu_{\perp})}{2\mu_1\mu_{\parallel}\epsilon_2-\mu_1^2-\mu_{\parallel}\mu_{\perp}}
\end{eqnarray}
\begin{eqnarray}
\vert H_{2z}\vert^2= \frac{\vert H_2\vert^2\mu_1(\mu_1\epsilon_2-\mu_{\perp} )}{\mu_1(\mu_1\epsilon_2-\mu_{\perp})-\mu_{\parallel}(\mu_1-\mu_{\parallel}\epsilon_2)}.
\end{eqnarray}
To obtain the relation for the electric field components, we employ Maxwell's equation
\begin{eqnarray}
  \vec{E} &=& \frac{i}{\omega\epsilon}\nabla\times \vec{H}
\end{eqnarray}
and derive the following relations
\begin{eqnarray}
E_{1y}=E_{2y}=\frac{\mu_0c^2}{\omega}(k_{1z}H_x-k_pH_{1z})
\end{eqnarray}
\begin{eqnarray}
\frac{\vert H_1\vert^2}{\vert H_2\vert^2}=\frac{\mu_{\parallel}(2\mu_1\mu_{\parallel}\epsilon_2-\mu_{\parallel}\mu_{\perp}-\mu_1^2)}{\mu_1[\mu_1(\mu_1\epsilon_2-\mu_{\perp})-\mu_{\parallel}(\mu_1-\mu_{\parallel}\epsilon_2)]}
\end{eqnarray}
\begin{eqnarray}
\vert E\vert^2=\vert E_{1y}\vert^2=\vert E_{2y}\vert^2= \mu_0^2c^2\vert H_1\vert^2\mu_1\frac{\mu_1^2-\mu_{\perp}\mu_{\parallel}}{2\mu_1\mu_{\parallel}\epsilon_2-\mu_{\perp}\mu_{\parallel}-\mu_1^2}.\nonumber\\
\end{eqnarray}

\subsection{Energy density}
The density of electromagnetic field energy in a dispersive medium is given by \cite{Landau}
\begin{eqnarray}
U=\frac{1}{2}\left[ \epsilon_0 \frac{\partial (\omega \epsilon)}{\partial \omega}E^2+\mu_0\frac{\partial (\omega \mu)}{\partial \omega}H^2\right].
\end{eqnarray}
Then in the upper space $z>0$, we have
\begin{eqnarray}
U_1&=&\frac{1}{4}(\epsilon_0\vert E_1\vert^2+\mu_0\mu_1\vert H_1\vert^2)e^{-2\vert k_{1z}\vert z}\nonumber \\
&=&\frac{1}{2}\mu_0\vert H_1\vert^2\frac{\mu_1\mu_{\parallel}(\mu_1\epsilon_2-\mu_{\perp})}{2\mu_1\mu_{\parallel}\epsilon_2-\mu_{\perp}\mu_{\parallel}-\mu_1^2}e^{-2\vert k_{1z}\vert z}.
\end{eqnarray}
In the superlattice $z<0$, the electromagnetic density is
\begin{widetext}
\begin{eqnarray}
U_2&=&\frac{1}{4}(\mu_0\mu_{\perp}\vert H_x\vert^2+\mu_0\mu_{\parallel}\vert H_{2z}\vert^2+\mu_0\omega\frac{\partial \mu_{\perp}}{\partial \omega}\vert H_x\vert^2+\mu_0\omega\frac{\partial \mu_{\parallel}}{\partial \omega}\vert H_{2z}\vert^2+\epsilon_2\epsilon_0\vert E\vert^2)e^{2\vert k_{2z}\vert z}\nonumber \\
&=&\frac{1}{4}\mu_0\vert H_2\vert^2[\frac{2\mu_1\mu_{\parallel}(\mu_1\epsilon_2-\mu_{\perp})}{\mu_1(\mu_1\epsilon_2-\mu_{\perp})-\mu_{\parallel}(\mu_1-\mu_{\parallel}\epsilon_2)}
-\frac{\omega\mu_{\parallel}(\mu_1-\mu_{\parallel}\epsilon_2)}{\mu_1(\mu_1\epsilon_2-\mu_{\perp})-\mu_{\parallel}(\mu_1-\mu_{\parallel}\epsilon_2)}\frac{\partial \mu_{\perp}}{\partial \omega}\nonumber \\
&&+\frac{\omega\mu_1(\mu_1\epsilon_2-\mu_{\perp})}{\mu_1(\mu_1\epsilon_2-\mu_{\perp})-\mu_{\parallel}(\mu_1-\mu_{\parallel}\epsilon_2)}\frac{\partial \mu_{\parallel}}{\partial \omega} ] e^{2\vert k_{2z}\vert z}.
\end{eqnarray}
\end{widetext}
The total energy density associated with the SPhPs is determined by integration over $z$ \cite{jpc-7-3547},
\begin{eqnarray}
\langle U_1 \rangle+\langle U_2 \rangle&=&\int_0^\infty U_1dz+\int^0_{-\infty}U_2dz\nonumber \\
&=&\frac{1}{8}\mu_0\frac{\vert H_2\vert^2}{\vert k_{2z}\vert}M(\mu_1,\mu_{\perp},\mu_{\parallel},\omega,\epsilon_2)\nonumber\\
&=&\frac{1}{8}\mu_0\frac{\vert H_1\vert^2}{\vert k_{1z}\vert}F(\mu_1,\mu_{\perp},\mu_{\parallel},\omega,\epsilon_2),
\end{eqnarray}
with
\begin{widetext}
\begin{eqnarray}
M(\mu_1,\mu_{\perp},\mu_{\parallel},\omega,\epsilon_2)&=&\frac{2\mu_1^2\mu_{\parallel}(\mu_1\epsilon_2-\mu_{\perp})-2\mu_{\parallel}^2\mu_{\perp}(\mu_1\epsilon_2-\mu_{\perp})}{\mu_1[\mu_1(\mu_1\epsilon_2-\mu_{\perp})
-\mu_{\parallel}(\mu_1-\mu_{\parallel}\epsilon_2)]}-\frac{\omega\mu_{\parallel}(\mu_1-\mu_{\parallel}\epsilon_2)}{\mu_1(\mu_1\epsilon_2-\mu_{\perp})-\mu_{\parallel}(\mu_1-\mu_{\parallel}\epsilon_2)}\frac{\partial \mu_{\perp}}{\partial \omega}\nonumber \\
&&+\frac{\omega\mu_1(\mu_1\epsilon_2-\mu_{\perp})}{\mu_1(\mu_1\epsilon_2-\mu_{\perp})-\mu_{\parallel}(\mu_1-\mu_{\parallel}\epsilon_2)}\frac{\partial \mu_{\parallel}}{\partial \omega},
\end{eqnarray}
\end{widetext}
and
\begin{widetext}
\begin{eqnarray}
F(\mu_1,\mu_{\perp},\mu_{\parallel},\omega,\epsilon_2)&=&\frac{2\mu_1\mu_{\perp}\mu_{\parallel}(\mu_1\epsilon_2-\mu_{\perp})-2\mu_1^3(\mu_1\epsilon_2-\mu_{\perp})}{\mu_{\perp}(2\mu_1\mu_{\parallel}\epsilon_2-\mu_{\perp}\mu_{\parallel}-\mu_1^2)}
+\frac{\omega\mu_1^2\mu_{\parallel}(\mu_1-\mu_{\parallel}\epsilon_2)}{\mu_{\perp}\mu_{\parallel}(2\mu_1\mu_{\parallel}\epsilon_2-\mu_{\perp}\mu_{\parallel}-\mu_1^2)}\frac{\partial \mu_{\perp}}{\partial \omega}\nonumber \\
&&-\frac{\omega\mu_1^3(\mu_1\epsilon_2-\mu_{\perp})}{\mu_{\perp}\mu_{\parallel}(2\mu_1\mu_{\parallel}\epsilon_2-\mu_{\perp}\mu_{\parallel}-\mu_1^2)}\frac{\partial \mu_{\parallel}}{\partial \omega}.
\end{eqnarray}
\end{widetext}

\subsection{Quantization of the surface fields}
So far we have treated the electric and magnetic fields with respect to surface phonon polariotons as classical variables.
The magnetic field in medium 1 is given by
\begin{eqnarray}
\vec{H}_1&=& A_{1k}\vec{u}_{1k}e^{ik_{p}x-i\omega t}e^{ik_{1z}z}+c.c
\end{eqnarray}
with
\begin{eqnarray}\label{L}
\vec{u}_{1k}=\frac{1}{\sqrt{\mathcal {L}}}(\vec{e}_{x}-\frac{k_p}{k_{1z}}\vec{e}_z).
\end{eqnarray}

$ A_{1k}$ is the amplitude that is related to the destruction operator for photons \cite{prb-82-035411}, and $\mathcal {L}$ has the
dimension of a length and will be fixed later  to normalize the energy of each mode \cite{prb-82-035411}.

The total energy of the surface waves is
\begin{eqnarray}\label{TE}
S(\langle U_1 \rangle+\langle U_2 \rangle)&=& S\frac{1}{8}\mu_0\frac{\vert H_1\vert^2}{\vert k_{1z}\vert}F\nonumber\\
&=&S\frac{1}{8}\mu_0 \frac{4\vert A_{1k}\vert^2}{\vert k_{1z}\vert}\frac{1}{\mathcal {L}}(1+\frac{\vert k_p\vert^2}{\vert k_{1z}\vert^2})F\nonumber\\
&=&2\mu_0S\vert A_{1k}\vert^2=\mu_0SA_{1k}^*A_{1k}+\mu_0SA_{1k}A_{1k}^*.\nonumber\\
\end{eqnarray}
We have used the degree of freedom to set $\mathcal {L}$ to simplify the above equation, i.e., we choose
\begin{eqnarray}
\mathcal {L}&=&\frac{1}{4}\frac{F}{\vert k_{1z}\vert}(1+\frac{\vert k_p\vert^2}{\vert k_{1z}\vert^2})\nonumber\\
&=&\frac{\mu_1(\mu_{\parallel}\epsilon_2-\mu_1)+\mu_{\parallel}(\mu_1\epsilon_2-\mu_{\perp})}{4\vert k_{1z}\vert\mu_1(\mu_{\parallel}\epsilon_2-\mu_1)}F.
\end{eqnarray}
Then we find that the expression for the surface wave energy (\ref{TE}) has the structure of the energy of a harmonic oscillator.
By taking the equivalence
\begin{eqnarray}
A_{1k}\rightarrow\sqrt{\frac{\hbar\omega(\vec{k})}{2\mu_0S}}\hat{a}_{\vec{k}}  \qquad A_{1k}^*\rightarrow\sqrt{\frac{\hbar\omega(\vec{k})}{2\mu_0S}}\hat{a}_{\vec{k}}^\dag
\end{eqnarray}
we can get the quantized Hamiltonian of the surface wave with mode vector $\vec{k}$
\begin{eqnarray}
\hat{H}&=&\frac{1}{2}\hbar\omega(\vec{k})\left(\hat{a}_{\vec{k}}^\dag\hat{a}_{\vec{k}}+\hat{a}_{\vec{k}}\hat{a}_{\vec{k}}^\dag\right).
\end{eqnarray}
The surface wave field is thus quantized by association of a quantum mechanical harmonic oscillator to each mode $\vec{k}$.
The operators $\hat{a}_{\vec{k}}$ and $\hat{a}_{\vec{k}}^\dag$ are annihilation and creation operators which destroy and create
a quantum of SPhPs with energy $\hbar \omega(\vec{k})$, and obey bosonic commutation relations $[\hat{a}_{\vec{k}},\hat{a}_{\vec{k'}}^\dag]=\delta_{\vec{k}\vec{k'}}$. A single quantized surface phonon polariton excitation is written as
$\vert 1\rangle_{k}=\hat{a}_{\vec{k}}^\dag\vert 0\rangle_k$, with $\vert 0\rangle_k$ the vacuum state of the system.
Furthermore, the field operator for the magnetic field in the upper space is given by
\begin{eqnarray}\label{B1}
\vec{B}_1&=& \sqrt{\frac{\hbar\omega}{2\mu_0S}}\mu_1\mu_0\hat{a}_{\vec{k}}\vec{u}_{1k}e^{ik_{p}x-i\omega t}e^{-\text{Im}(k_{1z})z}+\text{H.c.}\nonumber\\
\end{eqnarray}
This field operator will be used to investigate the magnetic coupling between NV spins and the SPhPs, which allows us
to provide a  quantum theory to describe the coupling between the NV spin ensemble and the quantized SPhPs.

Note that Ref. \cite{prb-71-125106} studies the classical theory,  while here we study the quantum theory of SPhPs. Also, Ref. \cite{prb-71-125106} focuses on bulk modes, while here we focus on surface modes. Other studies focus on piezoelectric superlattices, while here we focus on
piezomagnetic ones.

\subsection{Damping of the SPhPs}

We  consider the SPhP damping associated with the nonradiative loss to the crystal.  SPhPs decay nonradiatively due to  interacting
with the material in the form of phonon scattering, defect scattering, etc, which generally  depends
on the temperature and the composition of the crystals. The decay of the SPhP mode is frequency dependent,
and near the SPhP resonance frequency the SPhP decay is approximately equal to the damping constant of the
crystal \cite{jpc-7-3547}.
If the damping of the material is taken into account, without loss of generality we can add a damping term
to the piezomagnetic equations, in which case the
permeability function could be written in the form \cite{prb-69-085118,prb-77-075126}
\begin{eqnarray}
\mu_{\perp}(\omega)&=&\mu_{11}^s\frac{\omega_{\perp o}^2-\omega^2-i\kappa \omega}{\omega_{\perp L}^2-\omega^2-i\kappa \omega}.
\end{eqnarray}
For the surface phonon polaritons in the presence of damping,  a proper damping constant  between $\kappa\sim 0.001\omega_{\perp L}$ and $\kappa\sim 0.01\omega_{\perp L}$  can be chosen \cite{prb-69-085118,prb-77-075126}. Another useful figure of merit is the propagation length $L_\text{SPhP}$, which can be calculated from the decay time $\tau_\text{SPhP}\sim \kappa_\text{SPhP}^{-1}$ and group velocity $v_g$, ie., $L_\text{SPhP}=v_g/\kappa_\text{SPhP}$.  It is usually larger than the wavelength of SPhP modes in the low dissipation case \cite{apl-101-151109}.

\section{An ensemble of NV centers interacting with the quantized modes of SPhPs}

\subsection{A single NV spin interacting with a single SPhP mode}

The interaction
of a single NV center located at $\vec{r}_0$
with  the  total magnetic field can be written as
\begin{eqnarray}
\hat{H}_\text{NV}=\hbar D \hat{S}{_{z}^{2}}+\mu_Bg_s\ B_z\hat{S}{_{z}}+\mu_Bg_s \vec{B}(\vec{r}_0)\cdot\hat{\vec{S}}
\end{eqnarray}
with $g_s=2$ the Land\'{e} factor of the NV center,  $\mu_B$ the Bohr magneton,  and $\hat{\vec{S}}$ the spin operator of the NV center.
In the basis defined by the eigenstates of $\hat{S}_z$, i.e., $\{\vert m_s\rangle,m_s=0,\pm1\}$, with $\hat{S}_z\vert m_s\rangle=m_s\vert m_s\rangle$, we get
\begin{widetext}
\begin{eqnarray}
\hat{H}_\text{NV}&=&\sum_{m_s}\{ \langle m_s\vert[\hbar D \hat{S}_z^2+\mu_Bg_s \ B_z \hat{S}_z]\vert m_s\rangle\}\vert m_s\rangle \langle m_s\vert+\sum_{m_s,m'_s}\{ \langle m_s\vert\mu_Bg_s\vec{B}\cdot\hat{\vec{S}}\vert m'_s\rangle\}\vert m_s\rangle \langle m'_s\vert\nonumber\\
&=&\sum_{m_s}\{\hbar D m_s^2+\mu_Bg_s  B_z m_s\}\vert m_s\rangle \langle m_s\vert+\sum_{m_s,m'_s} \mu_Bg_s\hat{B}_x  \langle m_s\vert\hat{S}_x\vert m'_s\rangle\vert m_s\rangle \langle m'_s\vert
+\sum_{m_s } \mu_Bg_sm_s\hat{B}_z  \vert m_s\rangle \langle m_s\vert \nonumber\\
&=& (\hbar D +\mu_Bg_sB_z )\vert +1\rangle \langle +1\vert+ (\hbar D -\mu_Bg_sB_z )\vert -1\rangle \langle -1\vert+\mu_Bg_sB_{z0}(\vert +1\rangle \langle +1\vert-\vert -1\rangle \langle -1\vert)(\hat{a}_{\vec{k}}^\dag+\hat{a}_{\vec{k}})\nonumber\\
&&+\frac{\sqrt{2}}{2}\mu_Bg_sB_{x0} (\hat{a}_{\vec{k}}+\hat{a}_{\vec{k}}^\dag)(\vert 0\rangle \langle +1\vert+\vert +1\rangle \langle 0\vert)+\frac{\sqrt{2}}{2}\mu_Bg_sB_{x0} (\hat{a}_{\vec{k}}+\hat{a}_{\vec{k}}^\dag)(\vert 0\rangle \langle -1\vert+\vert -1\rangle \langle 0\vert)
\end{eqnarray}
\end{widetext}
Under the condition $|\Delta/2+D-\omega(\vec{k})|\ll\Delta/2$, with $\Delta=2\mu_Bg_sB_z/\hbar$, we can neglect the state $\vert m_s=-1\rangle$, due to the
external field moving it far out of resonance. The static magnetic field $B_z$ is about $3$ mT that can make the above assumptions valid.
This magnetic field is not a strong field, under which the lineal magnetic response of the system still holds. In this case, the static effect of the system is described by the static permeability $\mu^s_{11}$ and $\mu^s_{33}$. So the static magnetic field applied to split the NV spin states can be compatible with the  piezomagnetic superlattice, and will not affect the SPhP modes.
Then under the rotating-wave approximation  we can get the following Hamiltonian that describes the
interaction between a single NV spin and a  SPhP mode $\vec{k}$
\begin{eqnarray}
\hat{\mathcal{H}}&=&\frac{1}{2}\hbar \omega_0\hat{ \sigma}_{z}+{\hbar \omega(\vec{k})}\hat{a}_{\vec{k}}^\dag\hat{a}_{\vec{k}}\nonumber\\
&&+\frac{\hbar g_{\mu}(\vec{k},z_0)}{\sqrt{S}}  \hat{\sigma}_{+} \hat{a}_{\vec{k}}e^{ik_p x_0}+\text{H.c.}.
\end{eqnarray}

\subsection{An  NV spin ensemble interacting with the SPhP modes}

We now consider the interaction between an ensemble of  NV centers and the SPhP modes.
As depicted in Fig.~1(a),  an ensemble of  NV
centers is doped into a  diamond crystal of thickness $h$, and located at positions $\vec{r}_i$, each of which with a fixed quantization axis pointing
along one of the four possible crystallographic directions. If the orientations are equally distributed among the
four possibilities, and the external field is homogeneous, then a quarter of the NV spins can be made resonant with the SPhP mode.
In such a case, we  have the following Hamiltonian for $N$ NV spins in the resonant
subensemble interacting with the quantized surface mode $\vec{k}$
\begin{eqnarray}\label{HN}
\hat{\mathcal{H}}^N_{\vec{k}}&=& \sum_{i=1}^{N}\frac{1}{2}\hbar\omega_{i}\hat{ \sigma}_{z}^{i}+\hbar \omega(\vec{k})\hat{a}_{\vec{k}}^\dag\hat{a}_{\vec{k}}\nonumber\\
&&+\sum_{i=1}^{N}\frac{\hbar g_{\mu}(\vec{k},z_i)}{\sqrt{S}}(  \hat{\sigma}_{+}^{i} \hat{a}_{\vec{k}}e^{ik_p x_i}+\text{H.c.}),
\end{eqnarray}
where $\omega_{i}=\omega_{0}+\delta_{i}\simeq \omega_{0}$, and $\delta_{i}$  are random offsets accounting for the inhomogeneous broadening of the spin
ensemble.

We introduce the collective operators for the spin wave modes in the NV  ensemble
\begin{eqnarray}
\hat{S}_{\vec{k}}^\dag&=&\frac{1}{\sqrt{N}g^N_{\mu}(\vec{k})}\sum_{i=1}^{N}g_{\mu}(\vec{k},z_i)\hat{\sigma}_{+}^{i}e^{ik_p x_i}
\end{eqnarray}
with $g^N_{\mu}(\vec{k})=\sqrt{ \sum_{i=1}^{N}|g_{\mu}(\vec{k},z_i)|^2/N}$.
Consider the commutator $[\hat{S}_{\vec{k}}, \hat{S}_{\vec{k}'}^\dag]\equiv D(\vec{k}'-\vec{k})$ in the fully polarized limit:
\begin{eqnarray}
[\hat{S}_{\vec{k}}, \hat{S}_{\vec{k}'}^\dag]&=&\frac{1}{N}\sum_{i,i'}^{N}\frac{g_{\mu}(\vec{k}',z_i')g_{\mu}(\vec{k},z_i)}{(g^N_{\mu}(\vec{k}))^2}[\hat{\sigma}_{-}^{i},  \hat{\sigma}_{+}^{i'}] e^{-ik_p x_i}e^{ik'_p x_i'} \nonumber\\
&=&\frac{1}{N}\sum_{i}^{N}\frac{g_{\mu}^2}{(g^N_{\mu})^2}e^{i(k'_p -k_p) x_i}\sim \frac{1}{N}\int^{l/2}_{-l/2}e^{i(k_{p}'-k_{p})x}dx,\nonumber\\
\end{eqnarray}
with $l$ the extent of the sample along the $x$ direction. When $\Delta k=k_{p}'-k_{p}=2\pi/l$, the mode overlap $D(\vec{k}_j-\vec{k}_i)=0$, which means the spin wave modes  in the strongly polarized limit are orthogonal.
We finally get the
interaction Hamiltonian for the collective NV spin mode $\hat{S}_{\vec{k}}$ coupled to the SPhP mode $\hat{a}_{\vec{k}}$
\begin{eqnarray}
\hat{\mathcal{H}}^I_{\vec{k}}&=&  \hbar G_{\mu}^N(\vec{k})\left(  \hat{S}^\dag_{\vec{k}}\hat{a}_{\vec{k}} +\text{H.c.}\right).
\end{eqnarray}

\subsection{Decoherence of  the NV centers }

We now consider the decoherence of NV centers in a diamond crystal. In the case of an ensemble of NV centers, there will
be  magnetic dipole-dipole interactions with other spins like paramagnetic impurities in the diamond crystal,  resulting in large dephasing of the NV spins. The coupling of NV spins ($S_j$) with the surrounding impurity spins ($S_k$) is \cite{prl-105-210501}
\begin{eqnarray}
H_\text{spin}&=&\hbar \sum_{j,k}S_{j,z}\textbf{D}_{jk}\cdot \vec{e}_{k,z} S_{k,z}
\end{eqnarray}
The dipole interaction vector is given by
\begin{eqnarray}
\textbf{D}_{jk}=\frac{\mu_0g_s^2\mu_B^2}{4\pi\hbar}\frac{3(\vec{r}_{jk}\cdot \vec{e}_{z})\vec{r}_{jk}-\vec{e}_z}{r^3_{jk}}
\end{eqnarray}
with $\vec{e}_z$ the unit vector for the $z$ axis set by the NV crystal axis, and $\vec{r}_{jk}$   the distance vector between the two spins.
We can estimate the dephasing for a given spin bath. For a given nitrogen spin density $n_N$, the typical strength of the spin-spin interaction is about $\mu_0g_s^2\mu_B^2n_N/4\pi\hbar$.  This gives the typical dephasing rate of $\gamma_s\sim 2$ MHz for high nitrogen spin density of $n_N=10^{19}\text{cm}^{-3}$. Local strain and hyperfine interactions with nearby nuclear spins will also induce dephasing for the NV spins. The typical value of spin dephasing rate for this case is on the order of magnitude less than MHz. Current experiments demonstrate that the dephasing time for an NV spin ensemble is  in the microsecond range, with $\gamma_s/2\pi\sim 3$ MHz.

\subsection{The master equation }

The full dynamics of our
system that takes these incoherent processes into account
is described by the master equation \cite{prl-110-126801}
\begin{eqnarray}
\label{M1}
\frac{d\hat{\rho}(t)}{dt}&=&-\frac{i}{\hbar}[\hat{\mathcal{H}}^I_{\vec{k}},\hat{\rho}]+\gamma_\text{s}\mathcal{D}[\hat{S}_{\vec{k}}^\dag\hat{S}_{\vec{k}}]\hat{\rho}
+ \kappa_\text{SPhP}\mathcal{D}[\hat{a}_{\vec{k}}]\hat{\rho}\nonumber\\
\end{eqnarray}
with $\mathcal{D}[\hat{o}]\hat{\rho}=\hat{o}\hat{\rho}\hat{o}^\dag-\frac{1}{2}\hat{o}^\dag\hat{o}\hat{\rho}-\frac{1}{2}\hat{\rho}\hat{o}^\dag\hat{o}$
for a given operator $\hat{o}$.
We assume that
the sample is strongly polarized, which can be easily implemented by spin-selective optical pumping, and the
number of spin excitations is small compared to $N$.
In the low-excitation limit, the collective spin wave mode
$\hat{S}_{\vec{k}}$ behaves as bosons, i.e., magnons.
The lowest two
 magnon states are the state with all NV spins pointing
down, $\vert 0\rangle_{\text{Magn}}=\vert 0_1 0_2... 0_N\rangle$, and the state with a single  magnon
excitation $\vert1_k\rangle_{\text{Magn}}=\hat{S}_{\vec{k}}^\dag\vert 0\rangle_{\text{Magn}}$.
Then under the Hamiltonian (\ref{HN2}), the system exchanges energy coherently between
a quantum of the SPhP mode and a magnon
 before the decoherence processes dominate the interaction.

\subsection{Eigenfrequencies of the coupled magnon-polariton system }

The  coupling between the spin wave mode and the SPhP mode can be determined directly by
looking at the eigenfrequencies of the coupled system while the spin ensemble is tuned into resonance with the SPhP mode.
To obtain the system eigenvalues, we consider the non-Hermitian Hamiltonian
\begin{eqnarray}
  \hat{H}_{\text{n-H}} &=&  \hbar (\omega_k-i\gamma_s) \hat{S}_{\vec{k}}^\dag\hat{S}_{\vec{k}} +\hbar (\omega(\vec{k})-i\kappa_\text{SPhP})\hat{a}_{\vec{k}}^\dag\hat{a}_{\vec{k}}\nonumber\\
 &&+ \hbar G_{\mu}^N(\vec{k})(  \hat{S}^\dag_{\vec{k}}\hat{a}_{\vec{k}} +\text{H.c.}).
\end{eqnarray}
Using the non-Hermitian Hamiltonian where the decays are
taken into account, the eigenenergies and the broadenings
of the coupled system can be obtained as the real and imaginary parts of
the eigenvalues, respectively.
The eigenvalues of $\hat{H}_{\text{n-H}} $ are given by
 \begin{widetext}
 \begin{eqnarray}
   E^{\pm} &=& \hbar  \left[  \frac{\omega_k+\omega(\vec{k})-i(\gamma_s+\kappa_\text{SPhP})}{2} \pm [(G_{\mu}^N)^2+\frac{1}{4}(\omega_k-\omega(\vec{k})-i(\gamma_s+\kappa_\text{SPhP}))^2]^{1/2} \right].
 \end{eqnarray}
 \end{widetext}
Then we obtain the eigenfrequencies of the coupled magnon-polariton system
$\omega_{\pm}=\text{Re}(E^{\pm})/\hbar$.

%\bibliography{25}
%

\end{document}